\documentstyle[aps,preprint,eqsecnum,tighten]{revtex}

\begin{document}
\draft
\preprint{BARI - TH 129/93}
\date{\today}
\title{
UNSTABLE MODES IN \\ THREE-DIMENSIONAL SU(2) GAUGE THEORY }
\author{Paolo Cea$^{a,b}$ and Leonardo Cosmai$^b$}
\address{
$^a$Dipartimento di Fisica dell'Universit\`a di Bari,
70126 Bari, Italy\\
{\rm and}\\
$^b$Istituto Nazionale di Fisica Nucleare, Sezione di Bari,
70126 Bari, Italy\\
(E-mail: cea@bari.infn.it, cosmai@bari.infn.it)
}
\maketitle
\begin{abstract}
We investigate SU(2) gauge theory in a constant
chromomagnetic field in three dimensions both in the continuum
and on the lattice. Using a variational method to stabilize
the unstable modes, we evaluate the vacuum energy density
in the one-loop approximation.
We compare our theoretical results with the outcomes of
the numerical simulations.
\end{abstract}
\narrowtext

\section{INTRODUCTION}
\label{sec:introduction}

It was pointed out by different authors~\cite{Savvidy77,Pagels78}
several years
ago that for four-dimensional non~Abelian gauge theories without
matter fields in the one-loop approximation states with a constant
chromomagnetic field lie below the perturbative ground state.
This discovery stirred up a lot of interest. It was soon realized,
however, by Nielsen and Olesen~\cite{Nielsen78} that these states are
not stable due to long-range modes, the Nielsen-Olesen unstable
modes.
It was been, then, conjectured~\cite{Ambjorn79,Nielsen80} that the
vacuum field configurations which differ from the classical external
field only in the unstable mode sector could stabilize the
Nielsen-Olesen unstable modes.
The conclusion was reached that first the constant
chromomagnetic field would form a lattice, which, however, could be
unstable under quantum fluctuations. These quantum fluctuations
might result in the formation of a quantum liquid, the so-called
Copenhagen vacuum~\cite{Nielsen80}.

On the other hand, the problem of the unstable modes has been
reconsidered from a different point of view~\cite{Cea87}.
In Ref.~\cite{Cea87}, henceforth referred as I, it has been showed
that, by using variational techniques on a class of approximately
gauge-invariant  Gaussian wave functionals, the stabilization of
the Nielsen-Olesen modes contributes to the energy density with a
negative classical term which cancels the classical magnetic
energy.
Moreover the stabilization of the Nielsen-Olesen modes induces
a further background field which behaves non analytically in the
coupling constant and screens almost completely the external
chromomagnetic field.
As a consequence, even in the strong-field regime, which is the
naive perturbative regime, one deals with the non~perturbative regime.
This means that the calculation of the energy density even in the
one-loop approximation is non~perturbative~\cite{Maiani86}.
In other words the calculation of the vacuum energy is truly
non~perturbative and it mandates a completely non~perturbative
approach.

In a previous paper\cite{Cea91,Ambjorn89} we investigate four
dimensional U(1) and SU(2) lattice gauge theories in an external
constant background field. In particular, for the non~Abelian
case we looked for the effects of the unstable modes on the vacuum
energy density. However, the conclusion was reached that the
Monte~Carlo data did not display effects due to unstable modes
owing to finite size of the lattice. Indeed, in four
dimensions the unstable modes are long range. Then in lattice
simulations one must approximate the continuum and, at the same time,
work with a lattice big enough to allow unstable modes. So
the external fields which can be put on a lattice with periodic
boundary conditions have strength~\cite{Cea91,Damgaard88}:
\begin{equation}
\label{eq:1.1}
a^2 g B \simeq \frac{2 \pi}{L^2} n
\end{equation}
where $n$ is an integer, $L$ is the lattice size,
and $a$ the lattice spacing.\\
On general grounds~\cite{DiGiacomo91}, one expects that it exsists
a critical field $B_c$ above which the system becomes paramagnetic.
As a consequence, we need external magnetic fields lower than the
critical field $B_c$.
Using the estimation~\cite{DiGiacomo91} $B_c \sim T_c^2$,
$T_c / \Lambda_L \simeq 40$ and
\begin{equation}
\label{eq:1.2}
\Lambda_L a(\beta) = f(\beta) \simeq
\left( \frac{6 \pi^2 \beta}{11} \right)^{\frac{51}{121}}  \;
\exp\left\{ -\frac{3 \pi^2 \beta}{11} \right\}
\end{equation}
we obtain for the minimal field $B_{min}$ which can be put on
a lattice:
\begin{equation}
\label{eq:1.3}
\frac{B_{min}}{B_c} \simeq \frac{2 \pi}{L^2}
\frac{\sqrt{\beta}}{(40)^2} f^{-2}(\beta) \; .
\end{equation}
For instance, for $\beta=2.5$ one need $L > 12$ to have
$B_{min}/B_c < 1$. Therefore, we see that in four dimensions
working with external magnetic fields lower than the critical fields
requires sizeable lattices.
We can avoid these problems
by considering three-dimensional gauge theories. Indeed, the
three-dimensional SU(2) gauge theory is superrenormalizable.
Moreover, as we shall discuss later, the unstable modes
are not long range.

The purpose of this paper is twofold. First, in Sect.~II through
Sect.~V we consider the SU(2) gauge theory in a constant
chromomagnetic field in the fixed-time Schr\"odinger representation.
In Sect.~II we show that the quadratic part of the Hamiltonian
displays unstable modes. In Sect.~III we discuss the stabilization
of the unstable modes by using the variational procedure of I.
Section~IV is devoted to the calculation of the stabilized
vacuum energy density in the one-loop approximation.
In Sect.~V we analyze the Nielsen-Ninomiya Ansatz.

Second, in Sect.~VI we reanalyze the problem by employing
the lattice formulation of the gauge theories and
compare the numerical results with theoretical expectations.
Finally, our conclusions are drawn in Sect.~VII.

\section{SU(2) HAMILTONIAN IN A BACKGROUND FIELD}
\label{sec:SU(2)Hamiltonian}

In this section we consider the (2+1)~dimensional pure SU(2)
gauge theory in the temporal gauge. We shall follow closely
the method of~I. In the temporal gauge $A^a_0 =0$
the Hamiltonian is
\begin{equation}
\label{eq:2.1}
H = \frac{1}{2} \int d^2x  \,\,
\left\{
\left( E^a_i(\vec{x}) \right)^2 +
\left( B^a(\vec{x}) \right)^2
\right\}   \; ,
\end{equation}
where
\begin{equation}
\label{eq:2.2}
B^a(\vec{x}) = \frac{1}{2} \epsilon_{ij} F_{ij}^a (\vec{x}) \; ,
\end{equation}
and
\begin{equation}
\label{eq:2.3}
F_{ij}^a (\vec{x}) = \partial_i A^a_j(\vec{x}) -
\partial_j A^a_i(\vec{x}) +
g \epsilon^{abc} A^b_i(\vec{x}) A^c_i(\vec{x}) \;  .
\end{equation}
Note that in two spatial dimensions the chromomagnetic field
$B^a(\vec{x})$  is a (pseudo)scalar.

In the fixed-time Schr\"odinger representation the chromoelectric
field $E^a_i(\vec{x})$ acts as functional derivative
\begin{equation}
\label{eq:2.4}
E^a_i(\vec{x}) = + i \frac{\delta}{\delta A^a_i(\vec{x}) }
\end{equation}
on the physical states which are functionals obeying the
Gauss'~law:
\begin{equation}
\label{eq:2.5}
\left[ \partial_i \delta^{ab} + g \epsilon^{acb} A^c_i(\vec{x}) \right]
\frac{\delta}{\delta A^b_i(\vec{x}) } {\cal G}(A) = 0 \; .
\end{equation}
The effects of an external background field is incorporated by
writing
\begin{equation}
\label{eq:2.6}
A^a_i(\vec{x}) =  \bar{A}^a_i(\vec{x}) + \eta^a_i(\vec{x})
\end{equation}
where $\bar{A}^a_i(\vec{x})$ is the background field,
and $\eta^a_i(\vec{x})$ the fluctuating field.
We are interested in a constant abelian chromomagnetic field.
Thus we set
\begin{equation}
\label{eq:2.7}
\bar{A}^a_i(\vec{x}) =  \delta^{a3} \delta_{i2} x_1 B \;, \; \,
\vec{x} = (x_1, x_2) \; .
\end{equation}
It is straightforward to rewrite the Hamiltonian in terms
of the fluctuating fields. We get:
\begin{equation}
\label{eq:2.8}
H = V \frac{B^2}{2} + H^{(2)}+ H^{(3)}+ H^{(4)} \; ,
\end{equation}
where
\widetext
\begin{equation}
\label{eq:2.9}
H^{(2)} = \frac{1}{2} \int d^2x  \,\,
\left\{
- \frac{\delta^2}{\delta \eta^a_i(\vec{x}) \delta \eta^a_i(\vec{x})}
+ \eta^a_i(\vec{x}) D^{ab}_{ij}(\bar{A};\vec{x}) \eta^b_j(\vec{x})
- \left[ D^{ab}_{i}(\vec{x}) \eta^b_i(\vec{x}) \right]^2
\right\}   \; ,
\end{equation}
\narrowtext
\begin{equation}
\label{eq:2.10}
H^{(3)} = \frac{g}{2} \epsilon_{jk}
\epsilon_{j^{\prime} k^{\prime}}
\epsilon^{abc}
\int d^2x \,\,  \eta^b_j(\vec{x}) \eta^c_k(\vec{x})
D^{ad}_{j^{\prime}} (\vec{x}) \eta^d_{k^{\prime}}(\vec{x}) \; ,
\end{equation}
\begin{equation}
\label{eq:2.11}
H^{(4)} = \frac{g^2}{8} \epsilon_{jk}
\epsilon_{j^{\prime} k^{\prime}}
\epsilon^{abc} \epsilon^{a^{\prime}b^{\prime}c^{\prime}}
\int d^2x \,\,  \eta^b_j(\vec{x}) \eta^c_k(\vec{x})
\eta^{b^{\prime}}_{j^{\prime}}(\vec{x})
\eta^{c^{\prime}}_{k^{\prime}}(\vec{x}) \; .
\end{equation}
In Equations~(\ref{eq:2.8}-\ref{eq:2.11})
$D^{ab}_i(\vec{x})$ is the covariant derivative with respect
to the background field:
\begin{equation}
\label{eq:2.12}
D^{ab}_i(\vec{x}) = \partial_i \delta^{ab}
+ g \epsilon^{acb} \bar{A}^c_i (\vec{x}) \; .
\end{equation}
$D^{ab}_{ij}(\bar{A},\vec{x})$ is the operator
\begin{equation}
\label{eq:2.13}
D^{ab}_{ij}(\bar{A},\vec{x}) =
- \delta_{ij} D^{ac}_k(\vec{x}) D^{cb}_k(\vec{x})
+ 2 g B \epsilon_{ij} \epsilon^{3ab} \; .
\end{equation}
The Gauss'~constraint~(\ref{eq:2.5}) can be rewritten as
\begin{equation}
\label{eq:2.14}
\left[
D^{ab}_{i}(\vec{x}) + g \epsilon^{acb} \eta^c_i(\vec{x})
\right]
\frac{\delta \cal{G}(\eta)}{\delta \eta^b_i(\vec{x})}
= 0   \; .
\end{equation}
As is well known~\cite{Jackiw84}, the Gauss'~constraint ensures
that the physical states are invariant against
time-independent gauge transformations. It follows, then,
that the physical states are not normalizable. This
means that we must fix the residual gauge invariance.
Following I, we impose the covariant Coulomb constraint
\begin{equation}
\label{eq:2.15}
D^{ab}_{i}(\vec{x}) \eta^b_i(\vec{x}) = 0 \; .
\end{equation}
As discussed in I, the functional measure in the scalar product
between two physical states gets modified  by the Fadeev-Popov
determinant associated to the gauge-fixing~(\ref{eq:2.15}).\\
We are interested in the vacuum energy. The one-loop approximation
of the vacuum energy corresponds to consider the quadratic piece
of the Hamiltonian. From equation~(\ref{eq:2.8}) we get
\begin{equation}
\label{eq:2.16}
H_0 = \frac{V B^2}{2} + H^{(2)}    \; .
\end{equation}
In the same approximation the Gauss'~constraint reduces to
\begin{equation}
\label{eq:2.17}
D^{ab}_i(\vec{x}) \,
\frac{\delta \cal{G}(\eta)}{\delta \eta^b_i(\vec{x})} = 0
\end{equation}
Equation~(\ref{eq:2.17}) means that the physical states are
functionals of the fields transverse with respect to the
operator $D^{ab}_i(\vec{x})$.\\
To diagonalize $H^0$, it suffices to solve the eigenvalue equations
\begin{equation}
\label{eq:2.18}
D^{ab}_{ij}(\bar{A};\vec{x}) \,
\phi^b_j(\vec{x}) = \lambda \phi^a_i(\vec{x})
\end{equation}
with the conditions
\begin{equation}
\label{eq:2.19}
D^{ab}_{i}(\vec{x}) \,
\phi^b_i(\vec{x}) = 0   \; .
\end{equation}
The method to solve Eqs.~(\ref{eq:2.18}) and~(\ref{eq:2.19}) has
been discussed in detail in Appendix B of I.
So we merely write down the final results. We find the following
transverse eigenvectors:
\begin{eqnarray}
\label{eq:2.20}
&& \phi^a_i(N,\alpha=1;\vec{x}) = \delta^{a3} \,\,
\frac{e^{i \vec{k} \cdot \vec{x}}}{\sqrt{2 \pi}} \,
 \epsilon_i(\vec{k}) \; ,
\nonumber \\
&& N = \vec{k}, \quad \vec{k} \cdot \vec{\epsilon} = 0, \quad
\lambda(N, \alpha=1) = \vec{k}^2 \; .
\end{eqnarray}
\vspace{.25cm}
\widetext
\begin{eqnarray}
\label{eq:2.21}
&& \phi^a_j(N,\alpha=2;\vec{x}) =
\frac{1}{\sqrt{2}} \left(
\begin{array}{c}  1 \\ -i \\ 0 \end{array} \right) \,
\phi^{+}_j(N;\vec{x}) +
\frac{1}{\sqrt{2}} \left(
\begin{array}{c}  1 \\ +i \\ 0 \end{array} \right) \,
\phi^{-}_j(N;\vec{x})
\nonumber \\[.25cm]
&& N = (p,n) \quad \lambda(N,\alpha=2) = g B (2n+1), \quad n \ge 1
\end{eqnarray}
where
\begin{eqnarray}
\label{eq:2.22}
&& \phi^{+}_j(N,\alpha=2;\vec{x}) =
\frac{\alpha_1}{\sqrt{2}} \left(
\begin{array}{c}  1 \\ -i \\ 0 \end{array} \right) \,
F^{+}(p,n+1;\vec{x}) +
\frac{\alpha_2}{\sqrt{2}} \left(
\begin{array}{c}  1 \\ +i \\ 0 \end{array} \right) \,
F^{+}(p,n-1;\vec{x})
\nonumber \\[.25cm]
&& \phi^{-}_j(N,\alpha=2;\vec{x}) =
\left[ \phi^{+}_j(N,\alpha=2;\vec{x}) \right]^{*}  \; .
\end{eqnarray}
\narrowtext
In Equation~(\ref{eq:2.22}) $\alpha_1 = \sqrt{n} \, a$ and
$\alpha_2 = \sqrt{n+1} \,\,a$; the constant $a$ is fixed
by the normalization condition
\begin{equation}
\label{eq:2.23}
\int d^2 x \,\, \phi^{+}_i(N,\alpha;\vec{x}) \,
\phi^{-}_i(N^{\prime},\alpha;\vec{x}) = \delta_{N N^{\prime}}  \; .
\end{equation}
Moreover
\widetext
\begin{eqnarray}
\label{eq:2.24}
&& F^{+}(p,n;\vec{x}) =  \frac{e^{+ipx_2}}{\sqrt{2 \pi}}  \,
N_n \,\, H_n \left[ \sqrt{g} B
\left( x_1+\frac{p}{gB} \right) \right]
e^{-gB(x_1+\frac{p}{gB})^2} \; ,
\nonumber \\[.25cm]
&& N_n = \left( \frac{gB}{\pi} \right)^{\frac{1}{4}} \,
\left(2^n n! \right)^{-\frac{1}{2}}   \; .
\end{eqnarray}
Finally
\begin{eqnarray}
\label{eq:2.25}
&& \phi^a_j(N,\alpha=3;\vec{x}) =
\frac{1}{\sqrt{2}} \left(
\begin{array}{c}  1 \\ -i \\ 0 \end{array} \right)
\phi^{+}_j(N;\vec{x})
\, + \,
\frac{1}{\sqrt{2}} \left(
\begin{array}{c}  1 \\ +i \\ 0 \end{array} \right)
\phi^{-}_j(N;\vec{x})
\nonumber  \\[.25cm]
&& N=p,  \qquad \lambda(p,\alpha=3) = - gB    \; ,
\end{eqnarray}
\narrowtext
with
\begin{eqnarray}
\label{eq:2.26}
&& \phi^{+}_j(p;\vec{x}) =
\frac{1}{\sqrt{2}} \left(
\begin{array}{c}  1 \\ -i \\ 0 \end{array} \right) \,
F^{+}(p,n=0;\vec{x})
\nonumber   \\[.25cm]
&& \phi^{-}_j(p;\vec{x}) =
\left[ \phi^{+}_j(p;\vec{x})  \right]^{*}   \; .
\end{eqnarray}
The presence of the eigenvector $\phi^a_i(N,\alpha=3;\vec{x})$
with negative eigenvalue implies that the quadratic Hamiltonian
is not positive definite, and whence unbounded from below.
According we can decompose  the fluctuating field
into a stable fluctuation and an unstable one
\begin{equation}
\label{eq:2.27}
\eta^a_i(\vec{x}) =
\eta^a_{si}(\vec{x}) + \eta^a_{ui}(\vec{x})  \; ,
\end{equation}
where
\begin{eqnarray}
\label{eq:2.28}
&& \eta^a_{si}(\vec{x}) =
\sum_{N,\alpha=1,2} c(N,\alpha) \phi^a_i(N,\alpha;\vec{x})   \; ,
\nonumber   \\[.25cm]
&& \eta^a_{ui}(\vec{x}) =
\sum_{N} c(N,\alpha=3) \phi^a_i(N,\alpha=3;\vec{x})   \; .
\end{eqnarray}
Whereupon, the quadratic Hamiltonian reads
\begin{equation}
\label{eq:2.29}
H^{(2)} = H^{(2)}_s +  H^{(2)}_u
\end{equation}
with $H^{(2)}_s$ positive definite.
{}From Equation~(\ref{eq:2.29}) it follows that the vacuum functional
can be factorized as
\begin{equation}
\label{eq:2.30}
\cal{G}_0(\eta) = \cal{W}_0(\eta_s) \,\, \cal{Z}_0(\eta_u)   \; ,
\end{equation}
and the vacuum energy can be written as
\begin{equation}
\label{eq:2.31}
E(B) =  V \, \frac{B^2}{2} + E_s + E_u    \; .
\end{equation}
The contribution to the vacuum energy due to the stable modes is
readily obtained in the one loop approximation.\\
It is straightforward to show that the stable mode vacuum functional
is:
\begin{equation}
\label{eq:2.32}
\cal{W}_0(\eta_s) = \exp \left\{
-\frac{1}{4}  \int d^2x \, d^2y  \,\, \eta^a_{si}(\vec{x})
\left( G_s \right)^{ab}_{ij}(\vec{x},\vec{y})
\eta^b_{sj}(\vec{y})  \right\}     \; ,
\end{equation}
with
\begin{equation}
\label{eq:2.33}
\left( G_s \right)^{ab}_{ij}(\vec{x},\vec{y}) =
\sum_{N,\alpha=1,2} 2 \lambda^{\frac{1}{2}}(N;\alpha) \,
\phi^a_i(N,\alpha;\vec{x}) \, \phi^{b*}_j(N,\alpha;\vec{y})   \; .
\end{equation}
Then we obtain
\begin{equation}
\label{eq:2.34}
E_s = \frac{V}{2} \int \frac{d^2k}{\left( 2 \pi \right)^2}  \,\,
|\vec{k}|  \,\, + \,\,
V \, \frac{gB}{2 \pi} \sum^{\infty}_{n=1} \sqrt{gB(2n+1)}   \; .
\end{equation}
In obtaining Equation~(\ref{eq:2.34}) we used
\begin{equation}
\label{eq:2.35}
\int dp \,\, |F(p,n;\vec{x})|^2 \,\, = \,\, \frac{gB}{2 \pi}  \; .
\end{equation}
On the other hand, we have
\begin{equation}
\label{eq:2.36}
H^{(2)}_u = \frac{1}{2} \, \int d^2x \,\,
\left\{ - \frac{\delta^2}{\delta \eta^a_{ui}(\vec{x})
\delta \eta^a_{ui}(\vec{x})} -
gB \eta^a_{ui}(\vec{x})  \eta^a_{ui}(\vec{x})  \right\}   \; ,
\end{equation}
so that $H^{(2)}_u$ is unbounded from below.
In order to stabilize $H^{(2)}_u$  we must take into account
the quartic coupling which is of order of $g^2$.
This means that we must take care of the Gauss'~constraint
at least at the order $g^2$.

The main strategy of these calculation has been discussed
in Section III of I. In the next section we shall summarize
the main achievements of our analysis.

\section{STABILIZING THE UNSTABLE MODES}
\label{sec:Stabilizing}

According to the previous section, we would like to stabilize
the unstable modes by taking into account the quartic coupling.
To do this, we must construct the variational vacuum functional
which satisfies the Gauss'~constraint up to order of $g^2$.
We assume for the vacuum functional the form in
Eq.~(\ref{eq:2.30}). The presence in Eq.~(\ref{eq:2.36})
of a negative mass squared term suggests to try with a
shifted Gaussian for the functional $\cal{Z}_0(\eta)$:
\widetext
\begin{equation}
\label{eq:3.1}
\cal{Z}_0(\eta_0) =
\exp \left\{ -\frac{1}{4} \, \int d^2x \, d^2y \,\,
\left[ \eta^a_{ui}(\vec{x}) - u^a_i(\vec{x}) \right] \,
\left( G_u \right)^{ab}_{ij}(\vec{x},\vec{y}) \,
\left[ \eta^b_{uj}(\vec{y}) - u^b_j(\vec{y}) \right]
\right\}   \; ,
\end{equation}
\narrowtext
where
\begin{equation}
\label{eq:3.2}
\left( G_u \right)^{ab}_{ij}(\vec{x},\vec{y}) \, = \,
\sum_p 2 \rho(p) \phi^a_i(p,\alpha=3;\vec{x})
\phi^{b *}_j(p,\alpha=3;\vec{y})   \; ,
\end{equation}
\begin{equation}
\label{eq:3.3}
u^a_i(\vec{x}) = \sum_p b(p) \phi^a_i(p,\alpha=3;\vec{x})  \; .
\end{equation}
The vacuum functional $\cal{G}_0(\eta)$ satisfies the Gauss'~law
in the lowest approximation Eq.~(\ref{eq:2.17}). In order
to impose the full constraint Eq.~(\ref{eq:2.14}), we
write
\begin{equation}
\label{eq:3.4}
\cal{G}^{\prime}_0(\eta) =
\exp \left[ \Gamma_0(\eta) \right] \cal{G}_0(\eta)  \; ,
\end{equation}
and fix iteratively the functional $\Gamma_0(\eta)$.

To proceed, we evaluate the expectation value of the Hamiltonian
on the vacuum functional~(\ref{eq:3.4}). Fortunately
this rather long calculation parallels closely the one done
in the case of three spatial dimensions. The reader
interested in further details may consult section IV of I.

In the one-loop approximation, the stable mode contribution to
the vacuum energy has been already evaluated, Eq.~(\ref{eq:2.34}).
For the unstable modes we get in the same approximation
\widetext
\begin{eqnarray}
\label{eq:3.5}
E_u  =  && \frac{1}{2} \, \int d^2x \, d^2y \,\,
\delta(\vec{x}-\vec{y}) \, D^{ab}_{ij}(\bar{A};\vec{x}) \,
\left[ \left(G^{-1}_u \right)^{ab}_{ij}(\vec{x},\vec{y}) +
u^a_i(\vec{x}) u^b_j(\vec{y}) \right]
\nonumber \\[.25cm]
 && +  \frac{1}{8} \int d^2x \,\,
\left(G_u \right)^{aa}_{ii}(\vec{x},\vec{x})
\quad + \quad  E^{(4)}_u   \; ,
\end{eqnarray}
\begin{eqnarray}
\label{eq:3.6}
E^{(4)}_u  =  && \frac{g^2}{4} \, \epsilon_{jk}
\epsilon_{j^{\prime} k^{\prime}} \epsilon^{abc}
\epsilon^{a b^{\prime} c^{\prime} }  \,\,
\int d^2x \,\,
\left\{
u^b_j(\vec{x}) u^c_k(\vec{x})
u^{b^{\prime}}_{j^{\prime}}(\vec{x})
u^{c^{\prime}}_{k^{\prime}}(\vec{x}) \right.
\nonumber \\[.25cm]
&& + \left.
\left[
\left( G^{-1}_u \right)^{b b^{\prime}}_{j j^{\prime}}(\vec{x},\vec{x})
\,
u^c_k(\vec{x})  u^{c^{\prime}}_{k^{\prime}}(\vec{x}) \quad +
\quad \mbox{two permutations}  \,\, \right] \, \right\} \; .
\end{eqnarray}
\narrowtext
In order to minimize $E_u$ with respect to $\rho(p)$
and $b(p)$, we observe that from Eqs.~(\ref{eq:3.3}),
(\ref{eq:2.25}) and (\ref{eq:2.26}) it follows:
\begin{eqnarray}
\label{eq:3.7}
&& u^3_j(\vec{x}) = 0
\nonumber \\[.25cm]
&& u^{\pm}_j(\vec{x}) \,\, = \,\,
\frac{1}{\sqrt{2}} (u^1_j \pm i u^2_j)
\,\, =  \,\,
\frac{1}{\sqrt{2}} \,
\left( \begin{array}{c}  1 \\ \pm i \end{array} \right) \,
g^{\pm}(\vec{x})
\end{eqnarray}
with
\begin{equation}
\label{eq:3.8}
g^{\pm}(\vec{x}) = \sum_p b(p) \frac{e^{\pm i p x_2}}{\sqrt{2 \pi}}
\, \left( \frac{gB}{\pi} \right)^{\frac{1}{4}} \,
e^{-\frac{gB}{2} \left( x_1 + \frac{p}{gB} \right)^2}  \; .
\end{equation}
Using Equations~(\ref{eq:3.2}) and (\ref{eq:3.7}), we
recast $E_u$ into:
\widetext
\begin{eqnarray}
\label{eq:3.9}
E_u = &&  \frac{1}{2}
\sum_p \left[ \rho(p) - \frac{gB}{\rho(p)} \right]
- gB \int d^2x  \,\, g^{+}(\vec{x}) g^{-}(\vec{x})
\, + \,
\frac{g^2}{2} \int d^2x \,\,
\left[ g^{+}(\vec{x}) g^{-}(\vec{x}) \right]^2
\nonumber \\[.25cm]
&& + g^2 \sum_p \frac{1}{2 \rho(p)} \,
\int d^2x \,\,
\left\{ 2 \left| F^{+}(p,n=0;\vec{x}) \right|^2 \,
g^{+}(\vec{x}) g^{-}(\vec{x})  \right.
\nonumber \\[.25cm]
&& + \left.  \frac{1}{2} F^{+}(p,n=0;\vec{x}) g^{-}(\vec{x})
\quad + \quad
\frac{1}{2}  F^{-}(p,n=0;\vec{x}) g^{+}(\vec{x})  \right\}  \; .
\end{eqnarray}
\narrowtext
As discussed in I, the necessary condition
so that our  configuration contributes to the energy density is:
\begin{equation}
\label{eq:3.10}
\int d^2x \,\,  g^{+}(\vec{x})  g^{-}(\vec{x})
\propto V \equiv L^2    \;  .
\end{equation}
Moreover it is easy to show that
the minimum of the energy is attained for
\begin{equation}
\label{eq:3.11}
g^{+}(\vec{x}) g^{-}(\vec{x}) = K
\end{equation}
where $K$ is a positive constant. The condition~(\ref{eq:3.11})
is fulfilled with the choice:
\begin{equation}
\label{eq:3.12}
b(p) = \sqrt{2 \pi K} \, e^{i L p}
\end{equation}
in the thermodynamic limit $L \rightarrow \infty$~\cite{Ansatz5}.
{}From Equations~(\ref{eq:3.11}) and (\ref{eq:3.12}) we obtain
\begin{eqnarray}
\label{eq:3.13}
E_u = && \frac{1}{2}
\sum_p \left[ \rho(p)  - \frac{gB}{\rho(p)} \right]
- V K gB \, + \, \frac{g^2}{2} V K^2
\nonumber \\[.25cm]
&& + g^2 K \, \,
\sum_p \frac{1}{\rho(p)} \,
\int d^2x \,\, \left| F^{+}(p,n=0;\vec{x}) \right|^2   \; .
\end{eqnarray}
The last equation tells us that we can assume $\rho(p)$
independent on $p$. Thus, by taking into account that~\cite{Landau}
\begin{equation}
\label{eq:3.14}
\sum_p \, = \, V \, \frac{gB}{2 \pi}
\end{equation}
and
\begin{equation}
\label{eq:3.15}
\sum_p \left| F^{+}(p,n=0;\vec{x}) \right|^2
\, = \, \frac{gB}{2 \pi}     \; ,
\end{equation}
we get finally
\begin{equation}
\label{eq:3.16}
\frac{E_u}{V} \, = \, \frac{gB}{4 \pi}
\left[ \rho - \frac{gB}{\rho} \right] \, + \,
g^2 \frac{gB}{2 \pi} \frac{K}{\rho} \, - \,
gBK + \frac{g^2}{2} K^2    \; .
\end{equation}
Varying with respect to K, we get:
\begin{equation}
\label{eq:3.17}
K \, = \, \frac{gB}{g^2} \, - \,
\frac{gB}{2 \pi} \frac{1}{\rho}     \; .
\end{equation}
Inserting into Eq.~(\ref{eq:3.16}) and neglecting terms
of the same order as the two loop contributions,
we obtain the simple result:
\begin{equation}
\label{eq:3.18}
\frac{E_u}{V} \, = \, -\frac{B^2}{2} \, + \,
\frac{gB}{4 \pi} \left[ \rho + \frac{gB}{\rho} \right]   \; .
\end{equation}
Now, the minimization with respect to $\rho$ is straightforward:
\begin{equation}
\label{eq:3.19}
\rho^2 = gB    \; .
\end{equation}
Whence
\begin{equation}
\label{eq:3.20}
\frac{E_u}{V} \, = \, -\frac{B^2}{2} \, + \,
\frac{ \left( gB \right)^{3/2}}{2 \pi} \, + \,
\cal{O}\left( \frac{g^2}{2 \pi} \, \frac{gB}{2 \pi} \right)   \; .
\end{equation}
In conclusion we get for the total vacuum energy
in the one-loop approximation the remarkable result:
\begin{equation}
\label{eq:3.21}
\frac{E(B)}{V} \, = \, +\frac{B^2}{2} \, + \,
\frac{E_s}{V}  \, -\frac{B^2}{2}  \, + \,
\frac{\left( gB \right)^{3/2}}{2 \pi}  \, +  \,
\cal{O}\left( \frac{g^2}{2 \pi} \, \frac{gB}{2 \pi} \right)   \; .
\end{equation}
\begin{equation}
\label{eq:3.22}
\frac{E_s}{V} \, = \,
\frac{1}{2} \int \frac{d^2k}{\left( 2 \pi \right)^2} \,\,
\left| \vec{k} \right| \, + \,
\frac{gB}{2 \pi} \sum^{\infty}_{n=1} \sqrt{gB(2n+1)}    \; .
\end{equation}
The last two terms in Equation~(\ref{eq:3.21}) are the contributions
due to the stabilized unstable modes.
Note that, as in three spatial dimensions, the unstable modes
contribute to the vacuum energy density with a negative
classical term which cancels out the
classical magnetic energy term.

\section{THE STABILIZED VACUA}
\label{sec:StabilizedVacua}

In the previous section we evaluated the vacuum energy density
in the one-loop approximation. If we neglect the unstable mode
contribution our result should coincide with the real part of
the one-loop effective potential:
\begin{equation}
\label{eq:4.1}
V^{(2)}(B) = \frac{B^2}{2} \, + \, \frac{1}{2} \,
\int \frac{d^2k}{(2 \pi)^2}   \,\, |\vec{k}| \, + \,
\frac{gB}{2 \pi} \sum^{\infty}_{n=1} \sqrt{gB(2n+1)}  \; .
\end{equation}
Obviously $V^{(2)}(B)$ is divergent. However the theory is
superrenormalizable. This means that once we subtract the
free vacuum energy density we are left with a finite result.
As we show below this is the case and the final result
\widetext
\begin{equation}
\label{eq:4.2}
\Delta V^{(2)}(B) \equiv V^{(2)}(B) - V^{(2)}(0) \, = \,
\frac{B^2}{2} \, - \, \frac{(gB)^{3/2}}{2 \pi} \,
\left[ 1 - \frac{\sqrt{2}-1}{4 \pi}
\, \zeta \left( \frac{3}{2} \right)  \right]  \; ,
\end{equation}
\narrowtext
where $\zeta(z)$ is the Riemann's Zeta function, is in agreement
with the calculation by H. D. Trottier~\cite{Trottier91}.\\
Indeed, by using the identity:
\begin{equation}
\label{eq:4.3}
\sqrt{a} = -\int^{\infty}_0 \frac{ds}{\sqrt{\pi} \sqrt{s}} \,\,
\frac{d}{ds} e^{-as}  \; ,
\end{equation}
we have
\widetext
\begin{eqnarray}
\label{eq:4.4}
\frac{gB}{2 \pi} \sum^{\infty}_{n=1} \sqrt{gB(2n+1)} && =
-\frac{gB}{2 \pi} \sum^{\infty}_{n=1}
\int^{\infty}_0 \frac{ds}{\sqrt{\pi} \sqrt{s}} \,\,
\frac{d}{ds} e^{-gB(2n+1)s}
\nonumber \\[.25cm]
&& = -\frac{gB}{2 \pi}
\int^{\infty}_0 \frac{ds}{\sqrt{\pi} \sqrt{s}} \,\,
\frac{d}{ds}
\left[\frac{1}{e^{+gBs} - e^{-gBs}} - e^{-gBs} \right]   \; .
\end{eqnarray}
\narrowtext
Moreover, we observe that
\begin{equation}
\label{eq:4.5}
\int \frac{d^2k}{(2 \pi)^2} \,\, |\vec{k}| \, = \,
-\frac{1}{4 \pi} \int^{\infty}_0 \frac{ds}{\sqrt{\pi} \sqrt{s}} \,\,
\frac{d}{ds} \left( \frac{1}{s} \right)   \; .
\end{equation}
Thus, Equations~(\ref{eq:4.4}) and (\ref{eq:4.5}) ensure that
\begin{equation}
\label{eq:4.6}
\frac{3}{2} \int \frac{d^2k}{(2 \pi)^2} \,\, |\vec{k}| =
V^{(2)}(0)   \; .
\end{equation}
Whence
\widetext
\begin{equation}
\label{eq:4.7}
\Delta V^{(2)}(B) = \frac{B^2}{2} - \frac{(gB)^{\frac{3}{2}}}{2 \pi}
- \frac{gB}{2 \pi} \int^{\infty}_0 \frac{ds}{\sqrt{\pi} \sqrt{s}}
\,\,  \frac{d}{ds}
\left\{ \frac{1}{e^{+gBs} - e^{-gBs}} -
\frac{1}{2gBs}  \right\}    \; .
\end{equation}
\narrowtext
A change of the integration variable recasts Eq.~(\ref{eq:4.7})
into~\cite{Gradshteyn80}
\widetext
\begin{eqnarray}
\label{eq:4.8}
\Delta V^{(2)}(B) && =
\frac{B^2}{2} - \frac{(gB)^{\frac{3}{2}}}{2 \pi}
\left[ 1 + \int^{\infty}_0 \frac{dx}{\sqrt{\pi} \sqrt{x}} \,\,
\frac{d}{dx}
\left\{ \frac{1}{e^x - e^{-x}} - \frac{1}{2x} \right\} \right]
\nonumber \\[.25cm]
&& = \frac{B^2}{2} - \frac{(gB)^{\frac{3}{2}}}{2 \pi}
\left[1 - \sqrt{2}
\, \zeta \left( -\frac{1}{2},\frac{1}{2} \right) \right]
\end{eqnarray}
\narrowtext
where $\zeta(z,q)$ is the generalized Riemann's Zeta function.\\
Observing that~\cite{Gradshteyn80}
\begin{equation}
\label{eq:4.9}
\sqrt{2} \, \zeta \left( -\frac{1}{2},\frac{1}{2} \right) =
\frac{\sqrt{2} -1}{4 \pi} \, \zeta \left(  \frac{3}{2}  \right)  \; ,
\end{equation}
we obtain the promised result Eq.~(\ref{eq:4.2}).\\
The one-loop effective potential~(\ref{eq:4.8})
has a negative minumum
which, however, lies in a region where the one-loop approximation
is not trustworthy.\\
When we take into account
the unstable modes we get~\cite{energydensity}
\begin{equation}
\label{eq:4.10}
\frac{\Delta E(B)}{V} = + \frac{\sqrt{2} -1}{8 \pi^2}
\, \zeta\left( \frac{3}{2} \right)
\, (gB)^{3/2} \, + \,
\cal{O} \left( \frac{g^2}{2 \pi} \frac{gB}{2 \pi} \right)   \; ,
\end{equation}
In Equation~(\ref{eq:4.10}) the coefficient in front of
$(gB)^{3/2}$ is positive. As a consequence,
the minimum of the vacuum energy density is attained for $B=0$.

It should be stressed that the calculation of the full contribution
to the vacuum energy is non trivial even in the
one-loop approximation. As a matter of fact,
Equation~(\ref{eq:4.10}) does not include the contributions
to the energy arising from the interaction of the stable mode with
the induced background field $u^a_i(\vec{x})$ which behaves
non analytically in the coupling constant. Nevertheless, the
cancellation of the classical magnetic energy term due to the
stabilization of the unstable modes is sound. So that
we can write in general
\begin{equation}
\label{eq:4.11}
\frac{\Delta E(B)}{V} = a_{3/2} (gB)^{3/2} \, + \,
\cal{O} \left( \frac{g^2}{2 \pi} \frac{gB}{2 \pi} \right)   \; .
\end{equation}
In order to avoid the unphysical situation where the vacuum energy
density decreases without bound for increasing external magnetic
field, the constant $a_{3/2}$ should be positive, so that
the minimum is again $B=0$.\\
It is also interesting to note that the total background field
is now given by:
\begin{equation}
\label{eq:4.12}
\vec{A}^a_T(\vec{x}) =
\vec{\bar{A}}(\vec{x}) + \vec{u}^a(\vec{x})    \; .
\end{equation}
{}From Equations~(\ref{eq:3.7}), (\ref{eq:3.11}), and  (\ref{eq:3.17})
we obtain for the full field strength tensor
\begin{equation}
\label{eq:4.13}
F^3_{ij} = \frac{g}{2 \pi} \sqrt{gB}
\left( \delta_{i1} \delta_{j2} - \delta_{i2} \delta_{j1} \right)   \; .
\end{equation}
On the other hand the external field strength is
\begin{equation}
\label{eq:4.14}
F^{\mathrm{ext}}_{ij} \equiv
\partial_i \bar{A}^3_j - \partial_j \bar{A}^3_i =
B \left( \delta_{i1} \delta_{j2} - \delta_{i2} \delta_{j1} \right)  \; .
\end{equation}
It follows that
\begin{equation}
\label{eq:4.15}
\frac{F^3_{12}}{F^{\mathrm{ext}}_{12}} =
\frac{1}{2 \pi} \left( \frac{B}{g^3} \right)^{-\frac{1}{2}}   \; .
\end{equation}
Equation~(\ref{eq:4.15}) can be compared with the result
we should have obtained by neglecting the unstable modes:
\begin{equation}
\label{eq:4.16}
\frac{F^3_{12}}{F^{\mathrm{ext}}_{12}} =  1     \; .
\end{equation}
Equation~(\ref{eq:4.15}) tells us that the background field induced
by the stabilization of the unstable modes strongly screens
the external magnetic field.\\
In conclusions, the main results of this section are summarized
in Eqs.~(\ref{eq:4.12}) and (\ref{eq:4.15}).
Note, however, that our results
Eqs.~(\ref{eq:4.12}) and (\ref{eq:4.15}) are valid only in the
thermodynamic limit $L \rightarrow \infty$.
In view of the comparison with Monte Carlo simulations on a
finite lattice, it is worthwhile to discuss another Ansatz
due to H. B. Nielsen and N. Ninomiya~\cite{Ambjorn79} which
can be realized even on a finite lattice.

\section{THE NIELSEN AND NINOMIYA ANSATZ}
\label{sec:Nielsen}

In this Section we discuss the Ansatz by Nielsen and
Ninomiya~\cite{Ambjorn79}:
\begin{equation}
\label{eq:5.1}
g^{+}(\vec{x}) = \sum_p b(p) \, \frac{e^{ipx_2}}{\sqrt{2 \pi}} \,
\left( \frac{gB}{\pi} \right)^{1/4} \,
e^{-\frac{gB}{2} \left(x_1+\frac{p}{gB} \right)^2}
\end{equation}
\begin{equation}
\label{eq:5.2}
b(p) = \sqrt{2 \pi K_N} \, \sum_n \delta(p-nc) \;,  \qquad
c=\sqrt{2 \pi gB}
\end{equation}
with $K_N$ constant to be fixed by minimizing the vacuum energy
density. Using the definition of the third Jacobi $\theta$
function~\cite{Gradshteyn80}
\begin{equation}
\label{eq:5.3}
\theta_3(z,q) = 1 + 2 \sum_{n \ge 1} \, q^{n^2} \, \cos{2 n z} \; ,
\end{equation}
we rewrite Eq.(\ref{eq:5.1}) as
\begin{equation}
\label{eq:5.4}
g^{+}(\vec{x}) = \sqrt{K_N} \,
\left( \frac{gB}{\pi} \right)^{1/4} \,
e^{-\frac{gB}{2} x_1^2} \, \theta_3(z,q)
\end{equation}
where
\begin{equation}
\label{eq:5.5}
q= e^{- \pi}  \; ,
\end{equation}
\begin{equation}
\label{eq:5.6}
z= \frac{c}{2} \, (x_2 + i x_1)  \; .
\end{equation}
It is easy to show that
\widetext
\begin{eqnarray}
\label{eq:5.7}
\int d^2x \,\, g^{+}(\vec{x}) g^{-}(\vec{x})  &  = &
K_N \left( \frac{gB}{\pi} \right)^{1/2}\,
\int dx_1 \, dx_2 \,\, e^{-gBx_1^2} \,
|\theta_3(z,q)|^2
\nonumber \\[.25cm]
& = & K_N \left( \frac{gB}{\pi} \right)^{1/2} \,
\int dx_1 \, dx_2
\end{eqnarray}
\narrowtext
and
\begin{equation}
\label{eq:5.8}
\int d^2x \,\, \left[g^{+}(\vec{x}) g^{-}(\vec{x}) \right]^2 =
K_N^2 \, \frac{gB}{2 \pi} \, \left[ \theta_3(0,q) \right]^2
\, \int dx_1 \, dx_2    \; .
\end{equation}
Inserting into Eq.(\ref{eq:3.9}), we get
\widetext
\begin{eqnarray}
\label{eq:5.9}
E_u   & = &   \frac{1}{2} \,
\sum_p \left[ \rho(p) - \frac{gB}{\rho(p)} \right] -
gBK_N \, \left( \frac{gB}{2 \pi} \right)^{1/2} \, V +
\frac{g^2}{2} \, K_N^2 \, \left( \frac{gB}{2 \pi} \right) \,
\left[ \theta_3(0,q) \right]^2 \, V
\nonumber \\[.25cm]
& & + \,\,  g^2 \sum_p \frac{1}{\rho(p)} \,
\int d^2x \,\, |F^{+}(p,n=0;\vec{x})| \,
g^{+}(\vec{x})  g^{-}(\vec{x})   \;  .
\end{eqnarray}
\narrowtext
Assuming $\rho(p)$ independent on $p$,  and using
Eqs.~(\ref{eq:3.14}) and  (\ref{eq:3.15}),
we recast Eq.~(\ref{eq:5.9}) into:
\widetext
\begin{eqnarray}
\label{eq:5.10}
\frac{E_U}{V}   & = &  \frac{gB}{4 \pi}
\left[ \rho - \frac{gB}{\rho} \right] -
gB \, \left( \frac{gB}{2 \pi} \right)^{1/2} \, K_N
\, + \,
\frac{g^2}{2} \, \frac{gB}{2 \pi}  \,
K_N^2 \, \left[ \theta_3(0,q) \right]^2
\nonumber \\[.25cm]
& & + \,\, g^2 \, \frac{gB}{2 \pi} \,
K_N \, \left( \frac{gB}{2 \pi} \right)^{1/2}  \,
\frac{1}{\rho}   \; .
\end{eqnarray}
\narrowtext
Varying with respect to $K_N$ we obtain
\begin{equation}
\label{eq:5.11}
K_N \, \left( \frac{gB}{2 \pi} \right)^{1/2}  \, = \,
\frac{gB}{g^2[\theta_3(0,q)]^2} - \frac{gB}{2 \pi} \,
\frac{1}{[\theta_3(0,q)]^2} \, \frac{1}{\rho}   \; ,
\end{equation}
whereupon
\begin{equation}
\label{eq:5.12}
\frac{E_u}{V} = \frac{gB}{4 \pi} \,
\left[ \rho + \frac{g B \epsilon}{\rho} \right] -
\frac{B^2}{2[\theta_3(0,q)]^2}
\end{equation}
with
\begin{equation}
\label{eq:5.13}
\epsilon = \frac{2}{[\theta_3(0,q)]^2} - 1  \; .
\end{equation}
Note that $\theta_3(0,q) = 1 + 2 e^{-\pi} + 2 e^{-4 \pi} +
\cdots \simeq 1.0864$, so that $\epsilon>0$.\\
The minimization with respect to $\rho$ is now straightforward.
We get:
\begin{equation}
\label{eq:5.14}
\rho = \sqrt{g B \epsilon}   \; ,
\end{equation}
\begin{equation}
\label{eq:5.15}
\frac{E_u}{V} = - \frac{1}{2 [\theta_3(0,q) ]^2} \, B^2 \, + \,
\frac{\sqrt{\epsilon}}{2 \pi} (gB)^{3/2}  \; .
\end{equation}
Finally, the total vacuum energy density is
\widetext
\begin{eqnarray}
\label{eq:5.16}
\frac{\Delta E(B)}{V} & = &  \frac{1}{2} \,
\left[ 1 - \frac{1}{[\theta_3(0,q)]^2} \right] B^2 -
\frac{ (gB)^{3/2}}{2 \pi} \,
\left[ 1 - \sqrt{\epsilon} - \frac{\sqrt{2} -1}{4 \pi}
\zeta(\frac{3}{2}) \right]
\nonumber \\[.25cm]
& & + \,\, \cal{O} \left( \frac{g^2}{2 \pi} \,
\frac{gB}{2 \pi} \right)  \; .
\end{eqnarray}
\narrowtext
We stress once again that the coefficient of $(gB)^{3/2}$ in
Eq.~(\ref{eq:5.16}) does not include the contributions due to
the interaction between stable modes and the induced background
field $u^a_i(\vec{x})$.

Note that the coefficient of the
$(gB)^{3/2}$ term is negative, so that Eq.~(\ref{eq:5.16})
displays a negative minimum. However, the minimum lies
in a region where two-loop terms are sizeable.\\
It is instructive to evaluate the field-strength tensor.
The only non-zero component of $F^a_{\mu \nu}$ is
$F^3_{12}$. We obtain
\begin{equation}
\label{eq:5.17}
F^3_{12}(\vec{x}) = B - g K_N \,
\left( \frac{gB}{2 \pi} \right)^{1/2} \,
e^{-gBx_1^2} \, |\theta_3(z,q) |^2  \; ,
\end{equation}
one can show~\cite{Ambjorn79} that the chromomagnetic field forms
a square lattice in the $x_1 - x_2$ plane with lattice constant
$a_N=\sqrt{2 \pi/g B}$.\\
Defining a space averaging
\begin{equation}
\label{eq:5.18}
\left\langle F^3_{12}  \right\rangle =
\frac{ \int d^2x \,\, F^3_{12}(\vec{x})}{\int d^2x}   \; ,
\end{equation}
and using Eq.(\ref{eq:5.7}), we get
\begin{equation}
\label{eq:5.19}
\left\langle F^3_{12}  \right\rangle =
B - g K_N \,
\left( \frac{gB}{2 \pi} \right)^{1/2}   \; .
\end{equation}
Finally, Equations~(\ref{eq:5.11}) and (\ref{eq:5.14}) bring on
\begin{equation}
\label{eq:5.20}
\frac{\left\langle F^3_{12}  \right\rangle}{F^{\mathrm{ex}}_{12}} =
\left( 1 -  \frac{1}{[\theta_3(0,q)]^2} \right) +
\frac{1}{2 \pi \sqrt{\epsilon} [\theta_3(0,q)]^2} \,
\frac{1}{\sqrt{B/g^3}}   \; ,
\end{equation}
where  $F^{\mathrm{ex}}_{12}=B$.

The Nielsen and Ninomiya Ansatz is interesting in view of
the comparison with the lattice approach to be discussed
in the next section. Indeed the chromomagnetic lattice
could be more easily realized on the space-time lattice if the
two lattices are commensurate. In the lattice approach the
external chromomagnetic field is quantized due to the periodic
boundary conditions:
\begin{equation}
\label{eq:5.21}
a^2 gB = \frac{2 \pi}{L^2} n \; ,
\qquad  n \mbox{  integer}
\end{equation}
where $a$ is the space-time lattice spacing, and $L$ the
linear lattice
size in lattice units. From Equation~(\ref{eq:5.21})
we get
\begin{equation}
\label{eq:5.22}
\frac{a_N}{a} = \frac{L}{\sqrt{2n}}     \; .
\end{equation}
Thus the lattices are commensurate if $2n$ is the square of
an integer.

\section{BACKGROUND FIELDS ON THE LATTICE}
\label{sec:Background}

{}From the previous section we learned that the calculation of the
vacuum energy is truly non-perturbative even in the one-loop
approximation. Thus, we need a non-perturbative approach
which can be furnished by the lattice formulation of gauge theories.
In Ref.~\cite{Cea91} we studied the four-dimensional pure SU(2)
lattice gauge theory in an external magnetic field. We looked
for the effects due to unstable modes. Due to the limited size
of the lattice we used in the Monte Carlo simulations, we
failed in observing the unstable mode effects. The reason for
such a failure have been already discussed in
the Introduction.

The situation should improve considerably
working with three-dimensional SU(2) gauge theory.
Indeed in $d=3$ we can perform
Monte Carlo simulations on lattices of considerable size.
In addition, in two spatial dimensions the unstable modes
are not long range.\\
The pure SU(2) gauge theory is implemented on the lattice through
the standard Wilson action
\begin{equation}
\label{eq:6.1}
S_W = \beta \sum_{x, \mu > \nu} U_{\mu \nu}(x)
\end{equation}
where $U_{\mu \nu}(x)$ is the elementary plaquette in the
$(\mu, \nu)-$plane at the lattice site $x$.\\
The external background field on the lattice can be introduced
via an external current~\cite{Cea91,Damgaard88}. For the reader
convenience, we briefly summarize our method.\\
In the Euclidean continuum the background action reads
\begin{equation}
\label{eq:6.2}
S_B = \int d^3x \,\, j^a_{\mu}(x) A^a_{\mu}(x)   \; .
\end{equation}
Using the classical field equations:
\begin{equation}
\label{eq:6.3}
j^a_{\mu}(x) = D^{ab}_{\nu}(\bar{A}) \,
\bar{F}^b_{\nu \mu}(x) \; ,
\end{equation}
$\bar{F}^b_{\nu \mu}$ being the field strength tensor built from
the external fields $\bar{A}^a_{\mu}(x)$, and
\begin{equation}
\label{eq:6.4}
\bar{F}^a_{\mu \nu}(x) = \delta^{a3} \,
F^{\text{ext}}_{\mu \nu}(x)  \; ,
\end{equation}
we get from Eq.~(\ref{eq:6.2})
\begin{equation}
\label{eq:6.5}
S_B = -\frac{1}{2} \, \int d^3x \,\,
F^{\mathrm{ext}}_{\mu \nu}(x)
\left[ \partial_{\mu} A^3_{\nu}(x) -
\partial_{\nu} A^3_{\mu}(x)   \right]     \; .
\end{equation}
To discretize the action Eq.~(\ref{eq:6.5}), we must define the
Abelian-like piece
$\partial_{\mu} A^3_{\nu}(x) - \partial_{\nu} A^3_{\mu}(x)$
from the plaquette variables. To do this, we use
the so-called Abelian projection~\cite{tHooft81,Kronfeld87}.\\
To implement the Abelian projection, first we fix the gauge
by diagonalizing an operator $X(x)$ which transforms according to
the adjoint representation of the gauge group:
\begin{equation}
\label{eq:6.6}
V(x) \, X(x) \, V^{\dagger}(x) \,\, = \,\,
\mbox{diagonal matrix.}
\end{equation}
After that, we rewrite the gauge-fixed links
\begin{equation}
\label{eq:6.7}
\widetilde{U}_{\mu}(x) = V(x)U_{\mu}(x)V^{\dagger}(x+\mu)
\end{equation}
as
\begin{equation}
\label{eq:6.8}
\widetilde{U}_{\mu}(x) = W_{\mu}(x)U^A_{\mu}(x) \; ,
\end{equation}
where
\begin{equation}
\label{eq:6.9}
U^A_{\mu}(x) = \mbox{diag} \left[e^{i \theta^A_{\mu}(x)},
e^{-i \theta^A_{\mu}(x)} \right]
\end{equation}
\begin{equation}
\label{eq:6.10}
\theta^A_{\mu}(x) = \mbox{arg} \left\{ \left[
\widetilde{U}_{\mu}(x) \right]_{11} \right\}   \; .
\end{equation}
The Abelian-like links (\ref{eq:6.9}) are called the
Abelian projection of $U_{\mu}(x)$. The Abelian projected
plaquettes $U^A_{\mu\nu}$ are built form the Abelian projected
links in the usual manner. Clearly we have
\begin{equation}
\label{eq:6.11}
U^A_{\mu\nu}(x) = \mbox{diag} \left[e^{i \theta^A_{\mu\nu}(x)},
e^{-i \theta^A_{\mu\nu}(x)} \right]     \; .
\end{equation}
It is, thus, natural to define the Abelian field strength
tensor as
\begin{equation}
\label{eq:6.12}
F^A_{\mu\nu}(x) = \sqrt{\beta} \,\,
\mbox{tr} \left[ \frac{\sigma_3}{2i} \, U^A_{\mu\nu}(x) \right] =
\sqrt{\beta} \, \sin \theta^A_{\mu\nu}(x)  \; .
\end{equation}
As a consequence, we are led to consider the following background
action
\begin{equation}
\label{eq:6.13}
S_B = -\sqrt{\beta} \, \sum_x F^{\mathrm{ext}}_{\mu\nu}(x)
\, \sin{\theta^A_{\mu\nu}(x)}   \; .
\end{equation}
Taking into account the periodic boundary conditions of the lattice,
we write
\begin{equation}
\label{eq:6.14}
F^{\mathrm{ext}}_{\mu\nu}(x) =
\, \sqrt{\beta} \,
\sin{\theta^{\mathrm{ext}}_{\mu\nu}(x)}   \; ,
\end{equation}
\begin{equation}
\label{eq:6.15}
\theta^{\mathrm{ext}}_{\mu\nu}(x)  =
\frac{2 \pi}{L^2} \, n^{\mathrm{ext}}_{\mu\nu}(x)   \; ,
\end{equation}
the $n^{\mathrm{ext}}_{\mu\nu}(x)$'s being integers.
In Equation~(\ref{eq:6.15}), $L$ is the lattice size.\\
We are interested in a constant Abelian chromomagnetic field.
In this case the lattice action reads:
\begin{equation}
\label{eq:6.16}
S= +\beta \sum_{x,\mu > \nu} U_{\mu \nu}(x) -
\beta \sum_x \sin \theta^{\mathrm{ext}}_{12} \,
\sin \theta^A_{12}(x)   \; .
\end{equation}
The quantity of interest is the vacuum energy density at zero
temperature in presence of the external magnetic field.
In the naive continuum limit one can show easily that
the vacuum energy density is given by~\cite{nota21}
\begin{equation}
\label{eq:6.17}
E \left( F^{\mathrm{ext}}_{12}  \right) =
\beta \left[ P_s \left( F^{\mathrm{ext}}_{12}  \right) -
P_t \left( F^{\mathrm{ext}}_{12}  \right) \right]
\end{equation}
with
\begin{equation}
\label{eq:6.18}
P_s \left( F^{\mathrm{ext}}_{12}  \right) =
1 -\frac{1}{2} \,
\mbox{tr} \, U_{12} \left( F^{\mathrm{ext}}_{12}  \right)
\end{equation}
\begin{equation}
\label{eq:6.19}
P_t \left( F^{\mathrm{ext}}_{12}  \right) =
\sum_{i=1,2} \left[ 1 -\frac{1}{2} \,
\mbox{tr} \, U_{3i} \left( F^{\mathrm{ext}}_{12}  \right)
\right]    \; .
\end{equation}
We performed Monte Carlo simulations with the
action~(\ref{eq:6.16}) on lattices of size $L=20$, and $L=40$
in the weak coupling region $\beta \ge 7$. We measured the
quantity
\widetext
\begin{eqnarray}
\label{eq:6.20}
\Delta E \left( F^{\mathrm{ext}}_{12}  \right) & \equiv &
E \left( F^{\mathrm{ext}}_{12}  \right) - E(0) =
\nonumber \\[.25cm]
& =  & \beta \left\{
\left \langle P_s \left( F^{\mathrm{ext}}_{12}  \right) \right \rangle
\, - \,  \left \langle P_s \left( 0 \right) \right \rangle
\, + \,
\left \langle P_t \left( 0 \right) \right \rangle \, - \,
\left \langle P_t \left( F^{\mathrm{ext}}_{12}  \right) \right \rangle
\right\}   \; .
\end{eqnarray}
\narrowtext
Our simulations were done in the unfixed gauge, which corresponds to
set $V(x)= \openone$ in Eq.~({\ref{eq:6.7}).

We stress that the background action Eq.~(\ref{eq:6.13})
depends on the gauge-fixing. On the other hand, in the continuum
limit the energy density is gauge invariant. This means
that one should verify that the lattice definition~(\ref{eq:6.17})
does not depend on the gauge-fixing procedure.
In our previous study in four-dimensions we found
a very weak dependence on the gauge-fixing.
Moreover we are interested in the effects due to the unstable modes.
As we have already discussed, the unstable modes
modify in a dramatic way the vacuum energy density. So that the
effects we are looking for should manifest even without any
gauge fixing.

After discarding 500 sweeps, we collect 500 measurements (1 every
5 sweeps). Statistical errors were evaluated by the jackknife
algorithm.\\
In order to reduce the statistical errors the difference
$\langle P_s - P_t \rangle$ was evaluated directly during
Monte Carlo runs. Note that the vacuum energy difference can be
also measured directly by observing that, within statistical
errors, we have
\begin{equation}
\label{eq:6.21}
\left\langle \text{tr } U_{3i} \left( F^{\text{ext}}_{12} \right)
\right\rangle =
\left\langle \text{tr } U_{3i} (0) \right\rangle =
\left\langle \text{tr } U_{12} (0) \right\rangle   \; .
\end{equation}
In Figure 1 we display the adimensional energy density
$\Delta E \left(F^{\text{ext}}_{12} \right) /g^6$ versus the
adimensional external field strenght
$F^{\text{ext}}_{12}/g^3$ for four different values of $\beta$.

In general the vacuum energy can be affected by finite lattice
effects. To avoid lattice artefacts we used quite large lattices.
We found that for $20^3$ lattices there are no sizeable
finite size effects up to $\beta=10$ (compare full and open
circles in Fig.~1). On the other hand for $\beta>10$ we
need larger lattices (see full and open triangles in Fig.~1).
As a consequence for $\beta=12$ and $15$ we doubled the lattice
size, so that we feel that our numerical results are reliable.

In the continuum the adimensional energy density can depend on the
unique adimensional combination at our hands, namely $B/g^3$.
So we can write
\begin{equation}
\label{eq:6.22}
\frac{\Delta E(B)}{g^6} = f \left( \frac{B}{g^3} \right)  \; .
\end{equation}
In the one-loop approximation we have
\begin{equation}
\label{eq:6.23}
f(x) = a_{3/2} x^{3/2} + a_2 x^2
\end{equation}
where
\begin{equation}
\label{eq:6.24}
a_{3/2} = \frac{1}{2 \pi} \left[ 1 -
\frac{\sqrt{2} -1}{4 \pi} \, \zeta (3/2) \right]  \; ,
\qquad
a_2 = \frac{1}{2}
\end{equation}
if we neglect the unstable modes. Including the unstable modes
we obtained
\begin{equation}
\label{eq:6.25}
a_{3/2} = \frac{\sqrt{2} -1}{8 \pi^2} \, \zeta (3/2) \; ,
\qquad
a_2 = 0
\end{equation}
in the approximation of section~\ref{sec:StabilizedVacua},
and
\widetext
\begin{equation}
\label{eq:6.26}
a_{3/2} = -\frac{1}{2 \pi} \,
\left[ 1 - \sqrt{\epsilon} -
\frac{\sqrt{2} -1}{4 \pi} \, \zeta (3/2) \right] \; ,
\qquad
a_2 = \frac{1}{2} \, \left[ 1 -
\frac{1}{\left[ \theta_3(0,q) \right]^2} \right]
\end{equation}
\narrowtext
for the Nielsen and Ninomiya Ansatz. Note that the unstable modes
cause a drastic reduction of the classical magnetic energy term.\\
On the lattice there is another dimensional quantity,
namely the lattice spacing. Whereupon Equation~(\ref{eq:6.22})
is a non trivial check for the Monte Carlo outcomes. Any
deviations from the scaling law~(\ref{eq:6.22}) can be considered
as an indication of the granularity of the lattice.\\
A glance at Fig.~1 shows that the scaling law~(\ref{eq:6.22})
is satisfied quite well in the weak field strenght region
$F^{\text{ext}}_{12}/g^3 \leq 1$. But, for
strong field strenght $F^{\text{ext}}_{12}/g^3 > 1$ we observe
a small deviation from the scaling law~(\ref{eq:6.22}).
This found a quite natural explanation. Indeed, in order
to approximate the continuum, the magnetic length
$1/ \sqrt{g F^{\text{ext}}_{12}}$ should be greater than the
lattice spacing. Thus, at fixed lattice spacing, this condition gets
worse by increasing the external field strength.

In Figure~1 we compare our numerical results with the theoretical
expectations. The dotted line is the one-loop effective
potential~(\ref{eq:6.24}), the dashed line is Eq.~(\ref{eq:6.25}),
and the dot-dashed line Eq.~(\ref{eq:6.26}). Note that the
theoretical calculations are restricted
to the one-loop approximation. However this is not
a limitation at all. Indeed the two-loop terms are important for
\[
\frac{g^2}{2 \pi} \, \frac{gB}{2 \pi}  \,\, \gtrsim \,\,
\frac{(gB)^{3/2}}{2 \pi}   \; ,
\]
i.e.
\begin{equation}
\label{eq:6.27}
\frac{B}{g^3} \lesssim \frac{1}{4 \pi^2}    \; .
\end{equation}
Unfortunately our Monte Carlo simulations do not allow us
to appreciate the vacuum energy differences already for
$B/g^3 \lesssim 0.4 \,$.\\
{}From Figure~1 it is evident that the Monte Carlo data strongly
disagree with the one-loop effective potential Eq~(\ref{eq:6.24}).
Indeed the vacuum energy density is about an order of magnitude
smaller than the classical magnetic energy.
We feel that this reduction can be ascribed to the unstable modes.
As a matter of fact, our numerical results agree with
Eq.~(\ref{eq:6.25}) for
$F^{\text{ext}}_{12}/g^3  \lesssim 1 \,$. On the other hand for
$F^{\text{ext}}_{12}/g^3  \gtrsim 1$ the data can fitted by using
Eq.~(\ref{eq:6.23}) with $a_2  \sim  10^{-2}$.
However, as we have already stressed, the complete cancellation
of the classical magnetic energy can be attained for
configurations which satisfy Eq.~({\ref{eq:3.11}). This
condition can be fulfilled only in the thermodynamic limit.
When one deals with a finite lattice, it is natural to expect
the presence of a small classical-like term
in the vacuum energy density.

In order to have a further check, we looked at the Abelian
and non~Abelian chromomagnetic field strength. Indeed,
a further clear signature of the unstable modes resides
in a drastic reduction of the non~Abelian chromomagnetic field
with respect to the Abelian one.\\
In Figure~2 we display the ratio
$\left\langle F^{\text{A}}_{12} \right\rangle
/ F^{\text{ext}}_{12}$ versus
$F^{\text{ext}}_{12} / g^3$. From Figure~2 we see that the
system feels a rather strong Abelian chromomagnetic field.
Note, however, that the Abelian chromomagnetic field is
strongly affected by lattice artifacts.
In Ref.~\cite{Cea90} we found a similar situation
even in the U(1) lattice gauge theory.
The situation is quite different for non~Abelian
chromomagnetic field. As Figure~3 shows, the ratio
$\left\langle F^3_{12} \right\rangle
/ F^{\text{ext}}_{12}$ depends only on
$F^{\text{ext}}_{12} / g^3$ (the small deviations in the
strong field strength region have been already discussed).
Moreover the non~Abelian chromomagnetic field strength
is reduced by one order of magnitude with respect to the
Abelian one.
In Figure~3 we also compare the numerical data with
Eqs.~(\ref{eq:4.15}) and (\ref{eq:5.20}).
A few comments are in order.
Without  unstable modes and in the one-loop approximation
we should have obtained
\begin{equation}
\label{eq:6.28}
\frac{ \left\langle F^3_{12} \right\rangle}
{F^{\text{ext}}_{12}} = 1      \; .
\end{equation}
Equation~(\ref{eq:6.28}) is the dotted line in Fig.~3.
We see that Eq.~(\ref{eq:6.28}) is in total disagreement with
the numerical results. In addition, if we take into account that
a small quadratic term in the energy density adds a constant
term of the same order in Eq.~(\ref{eq:4.15}), then we can
conclude that the data corroborate Eq.~(\ref{eq:4.15})
at least in the region
$F^{\text{est}}_{12} / g^3 \gtrsim 0.5 \,$.
As concerns the weak field strength region, on the one hand
we expect that in Eq.~(\ref{eq:4.15}) there are sizeable
effects due to higher order contributions. On the other hand,
the numerical results are not reliable because the magnetic length
became comparable to the lattice size.\\
Thus we are confident that Figs.~1 and 3 together give evidence
of the unstable modes on the lattice.

Let us conclude this section by comparing our approach to
the one recently proposed by
H. D. Trottier and R. M. Voloshyn~\cite{Trottier92}. These authors
consider three-dimensional lattice gauge theories in a
background field. The background field is induced by means of an
external current~Eq.(\ref{eq:6.2}). Thus in the continuum
the background action they used coincides with Eq.~(\ref{eq:6.5}).
For the U(1) gauge theory the action of Ref.~\cite{Trottier92}
agrees with our lattice background action Eq.~(\ref{eq:6.13}).
However the authors of Ref.~\cite{Trottier92} do not enforce
periodic boundary conditions, so that the external background field
strength is not restricted by
the constraint Eq.~(\ref{eq:6.15})~\cite{nota25}.
As concerns the SU(2) gauge theory, Trottier and Woloshyn adopt a
different discretization of the Abelian field strength tensor:
\begin{equation}
\label{eq:6.29}
F^A_{\mu \nu} (x) = \sqrt{\beta}  \,\,  \text{tr }
\left\{ \frac{\sigma_3}{2 i} \, \left(
U_{\mu \nu} - \left[ U_{\mu}, U_{\nu} \right] \right] \right\}  \; .
\end{equation}
It should be stressed that, whatever discretization one
chooses the total action is no longer SU(2)-invariant, but
only U(1)-invariant. This means there is always the freedom
of reducing the original local SU(2) invarance to a local
U(1) invariance with a suitable gauge fixing.
In particular, the authors of Ref.~\cite{Trottier92} used
the discretization Eq.~(\ref{eq:6.29}) without gauge-fixing.
As concerns the numerical results,
the authors of Ref.~\cite{Trottier92} found that the
vacuum energy density is more than an order of magnitude smaller
than the classical magnetic energy over the whole range of
the applied external magnetic field, in qualitative agreement
with our results.\\
However, for $F^{\text{ext}}_{12} / g^3 \lesssim 1$
the vacuum energy density of Ref.~\cite{Trottier92} is negative.
This result looks puzzling to us because we do not expect that
in the weak external field strength region the vacuum energy
density displays a dramatic dependence on the discretization
of the Abelian field strenght tensor. To check this point,
we performed Monte Carlo simulations for
$F^{\text{ext}}_{12} / g^3 \le 1$ by using Eq.~(\ref{eq:6.29})
in the background action. It turns out that
the vacuum energy density difference is
\underline{positive} even though it is about a factor two
smaller than our previous results.
This disagreement can be ascribed to the different discretization
adopted for the Abelian field strength.
It turns out, however, that
the discretization Eq.~(\ref{eq:6.29}) leads to
a vacuum energy density which is close to Eq.~(\ref{eq:6.25}).
Thus the method of Ref.~\cite{Trottier92} corroborate the evidence
of unstable modes on the lattice.

Contrarily to our findings
the authors of Ref.~\cite{Trottier92}
found that the vacuum energy density is negative for
$F^{\text{ext}}_{12}/g^3 \lesssim 1 \,$.
We do not believe that this disagreement can be ascribed to
the different boundary conditions. We do not yet understand
the reasons of this discrepancy. We hope to clarify this
point in future studies.

\section{CONCLUSIONS}
\label{sec:conclusions}

Let us conclude by summarizing the main achievements of this paper.
We investigated the three-dimensional SU(2) gauge theory
in a constant chromomagnetic field, both in the continuum
and on the lattice. As in the four-dimensional case, we found
the presence of unstable modes. The stabilization procedure of
the unstable modes introduces a new background field with
non~analytic behaviour in the coupling constant. The most striking
consequences of the induced background field are the cancellation
(or a drastic reduction) of the classical magnetic energy, and
an almost complete screening of the applied magnetic field.
Moreover, we found that the non~analytic behaviour of the induced
background field makes the calculation of the ground-state energy
highly non~trivial. We studied, then, the problem with the
non~perturbative techniques offered us by the lattice approach to
the gauge theories. We feel that we have convincingly put out
the evidence of the unstable modes.

We would like to end
by briefly discussing some consequences of our finding.
As is known since long time, in the non~Abelian gauge theories
in the perturbative one-loop approximation, states with a constant
chromomagnetic field lie below the perturbative ground state.
It follows that these states could be a better approximation
to the true ground state than the perturbative one. In particular,
it was believed~\cite{Nielsen80} that from these states one could
set up an approximate ground state which confines color charges.
However, the presence of unstable modes leads to a muddled state.
In the case of (2+1)-dimensions, we argued
(see the discussion after
Eq.~(\ref{eq:4.10})) that, after the stabilization of the unstable
modes, the states with a constant chromomagnetic background
field are not energetically favoured with respect to the the
perturbative ground state. This means that these states
are not relevant to the confinement at least in three spacee-time
dimensions. However, if we belive that the confinement mechanism
does not depend on the space-time dimensions, then we are led
to the same conclusions even in
the more interesting four-dimensional case.


%


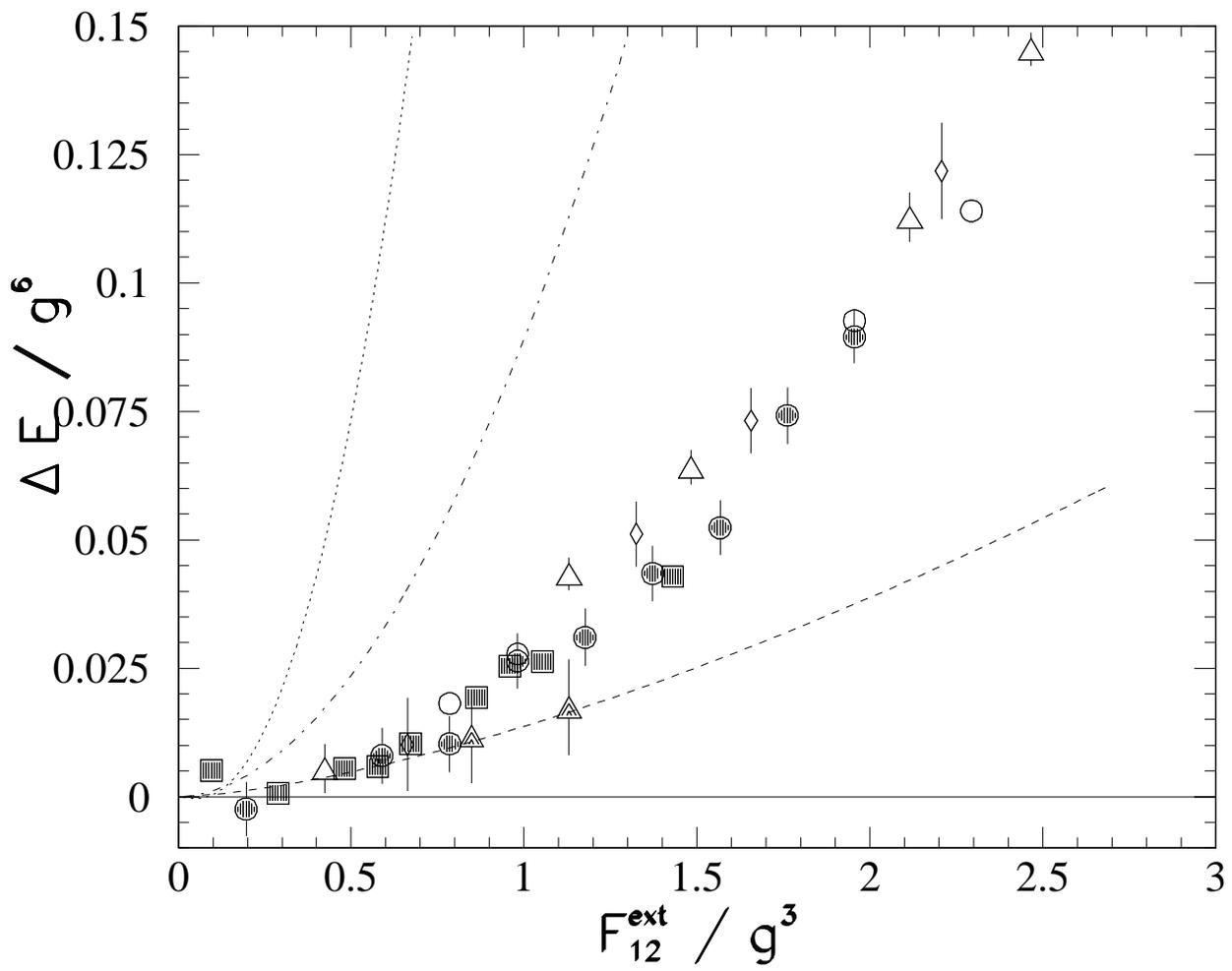
\begin{figure}
\protect\caption{Vacuum energy density versus
the applied magnetic field.
Full and open symbols refer to L=20 and L=40 respectively.
Squares correspond to $\beta=7$, circles to $\beta=10$,
triangles to $\beta=12$, and diamonds to $\beta=15$.
The curves are discussed in the text.}
\end{figure}

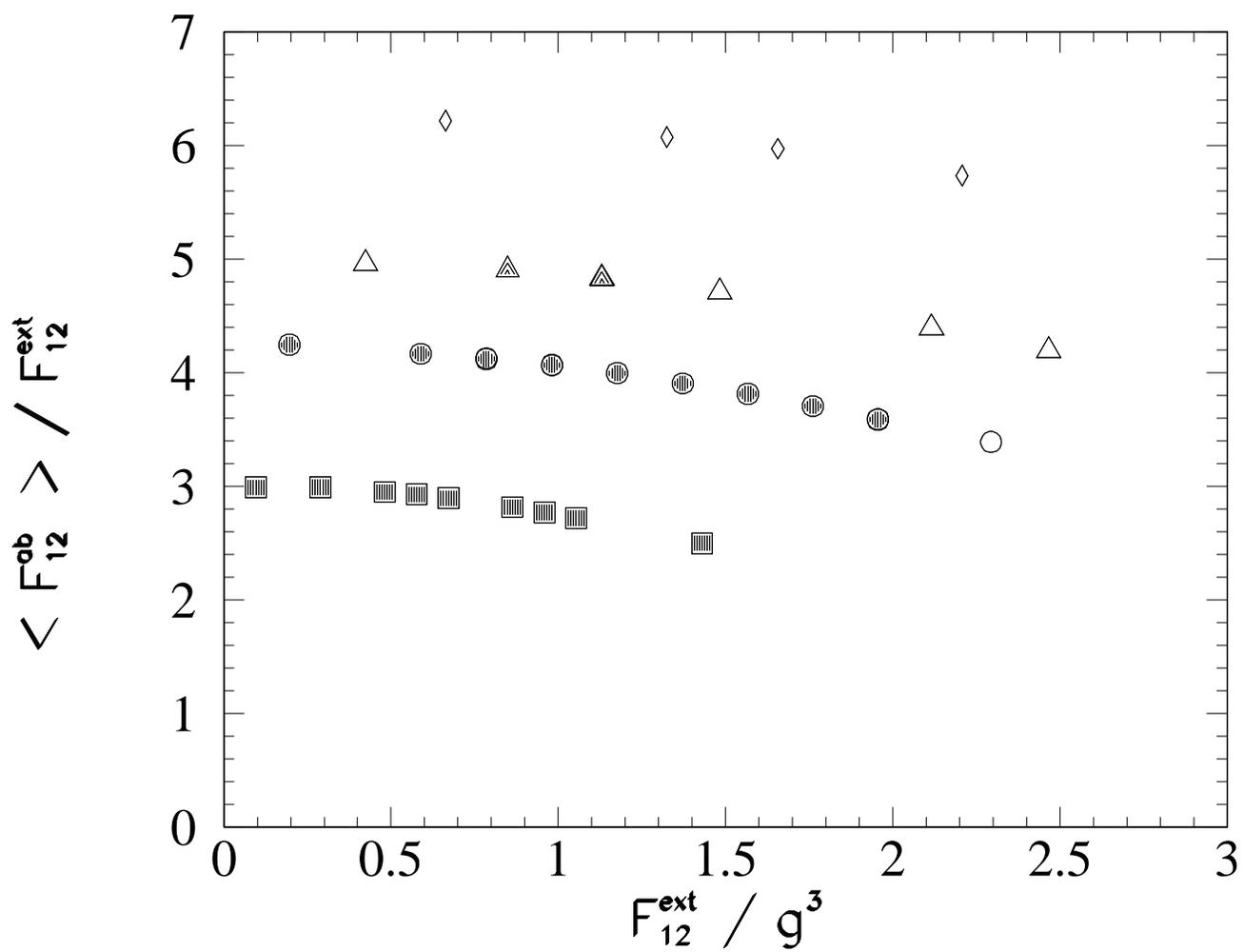
\begin{figure}
\protect\caption{The Abelian chromomagnetic field strength
versus the applied magnetic field
(symbols as in Fig.~1).}
\end{figure}

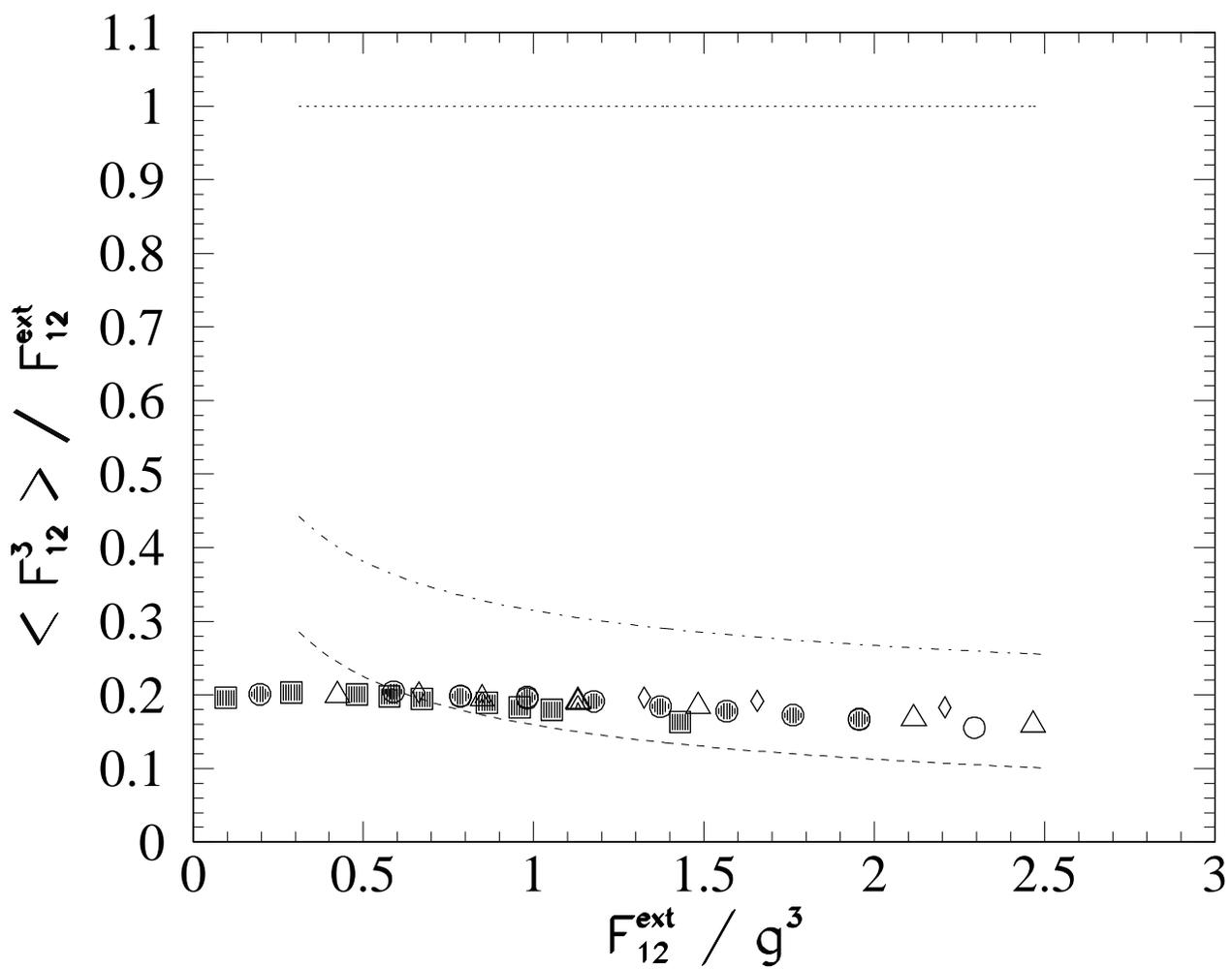
\begin{figure}
\protect\caption{The full chromomagnetic field strength versus
the applied magnetic field (symbols as in Fig.~1). The
dashed line is Eq.~(\protect\ref{eq:4.15}), the dot-dashed line
Eq.~(\protect\ref{eq:5.20}), and the dotted line is
Eq.~(\protect\ref{eq:6.28}).}
\end{figure}


\begin{references}
\bibitem{Savvidy77}G. K. Savvidy, Phys. Lett. {\bf 71B}, 133 (1977);\\
S. G. Matinyan and G. K. Savvidy, Nucl Phys. {\bf B134}, 539 (1978).
\bibitem{Pagels78} H. Pagels, {\it Lectures at Coral Gables}, Florida,
1978.
\bibitem{Nielsen78}N. K. Nielsen and P. Olesen, Phys. Lett. {\bf 79B},
304 (1978);\\
N. K. Nielsen and P. Olesen, Nucl. Phys. {\bf B144}, 376 (1978).
\bibitem{Ambjorn79} J. Ambjorn, N. K. Nielsen and P. Olesen,
Nucl. Phys. {\bf B152}, 75 (1979);\\
H. B. Nielsen and N. Ninomiya, Nucl. Phys. {\bf B156}, 1 (1979);\\
J. Ambjorn and P. Olesen, Nucl. Phys. {\bf B170}, 60 (1980);
{\it ibid.} 265.
\bibitem{Nielsen80} For a review see:
H. B. Nielsen, {\em Confinement with special
emphasis on the Copenhagen vacuum}, in Particle Physics (1980),
eds. I. Andric, I. Dadic and N. Zovko
(North-Holland, Amsterdam, 1981).
\bibitem{Cea87} P. Cea, Phys. Lett. {\bf B193}, 268 (1987);\\
P. Cea, Phys. Rev. {\bf D37}, 1637 (1988),  and references therein.
\bibitem{Maiani86} A similar conclusion was reached in
L. Maiani, G. Martinelli, G. C. Rossi, and
M. Testa, Nucl. Phys. {\bf B273}, 268 (1986).
\bibitem{Cea91} P. Cea and L. Cosmai, Phys. Lett. {\bf B264}, 415 (1991).
\bibitem{Ambjorn89} For a different approach see:
J. Ambjorn, V. K. Mitrjushkin, V. G. Bornyakov,
and A. M. Zadorozhny, Phys. Lett. {\bf B225}, 153 (1989).
\bibitem{Damgaard88} P. H. Damgaard and U. M. Heller,
Phys. Rev. Lett. {\bf 60}, 1246 (1988);
Nucl. Phys. {\bf B309}, 625 (1988);
Nucl. Phys. {\bf B324}, 532 (1989).
\bibitem{DiGiacomo91} A. Di Giacomo, invited talk at {\em
Convegno di Meccanica Statistica e Teoria dei Campi non Perturbativa},
March 25-28, 1991, Bari, Italy;\\
L. Del Debbio, A. Di Giacomo, M. Maggiore, and
\v{S}. Olejn\'{\i}k, Phys. Lett. {\bf B267}, 254 (1991).
\bibitem{Jackiw84} See for instance:
R. Jackiw, in {\em Relativity, Groups and
Topology~II}, Proceedings of Les~Houches Summer School, edited
by B. S. De~Witt and R. Stora (North-Holland, Amsterdam, 1984).
\bibitem{Ansatz5} We discuss another successfull Ansatz
in Section 5.
\bibitem{Landau} See, for instance: L.D. Landau,
{\em Mechanique Quantique},
(Editions Mir, Moscow, 1966) p.498.
\bibitem{Trottier91} H. D. Trottier, Phys. Rev. {\bf D44}, 464 (1991).
\bibitem{Gradshteyn80} I. S. Gradshteyn and I. M. Ryshik,
{\em Table of Integrals, Series, and Products},
(Academic Press, 1980).
\bibitem{energydensity} Note that in this case the energy density
does not coincide with the one-loop effective potential.
Indeed the stabilization of the unstable modes, going beyond
the perturbation theory, does not allow to define
the perturbative effective potential.
\bibitem{tHooft81} G. 't~Hooft, Nucl. Phys. {\bf B190}, 455 (1981).
\bibitem{Kronfeld87} A. S. Kronfeld, G. Schierholz, and
U.-J. Wiese, Nucl. Phys. {\bf B293}, 461 (1987).
\bibitem{nota21} In this section $E$ indicates the vacuum energy
density.
\bibitem{Cea90} P. Cea and L. Cosmai, Phys. Lett. {\bf B249},
114 (1990).
\bibitem{Trottier92} H. D. Trottier and R. M. Voloshyn,
{\em The Savvidy ferromagnetic vacuum in three-dimensional
lattice gauge theory}, TRIUMF TRI-PP-92-98, October 1992, Preprint.
\bibitem{nota25} This difference does not matter
in the present case. Indeed we checked that,
by using the periodic
lattice, the Monte Carlo simulations gave results which agree
with Ref.~{\protect\cite{Trottier92}},
namely the vacuum energy
density was close to the classical value
$\frac{1}{2} \left( F^{\text{ext}}_{12} \right)^2$.
\end{references}
\end{document}